\documentclass{PoS}
\usepackage{overpic}
\usepackage{amsmath}
\usepackage{amstext}
\usepackage{relsize}
\usepackage{booktabs}
\usepackage{multirow}
\usepackage{xspace}

\input{babarsym}

\newcommand{\Vusvalue} {\ensuremath{0.2255 \pm 0.0024}\xspace}

\newcommand{\gmgevalue} {\ensuremath{1.0036 \pm 0.0020}\xspace}

\newcommand{\gtgmpvalue} {\ensuremath{0.986 \pm 0.006}\xspace}
\newcommand{\gtgmkvalue} {\ensuremath{0.983 \pm 0.009}\xspace}
\newcommand{\gtgmhvalue} {\ensuremath{0.985 \pm 0.005}\xspace}

\newcommand{\tauknu}   {\ensuremath{ \tau^{-} \to K^{-} \nut}\xspace}
\newcommand{\taupinu}   {\ensuremath{ \tau^{-} \to \pi^{-} \nut}\xspace}
\newcommand{\tauhnunc}   {\ensuremath{ \tau \to h \nut}\xspace}

\newcommand{\tauenu}   {\ensuremath{ \tau^{-} \to e^{-} \nueb \nut}\xspace}
\newcommand{\taumunu}  {\ensuremath{ \tau^{-} \to \mu^{-} \numb \nut}\xspace}

\newcommand{\BFtautoknu}    {\ensuremath{\BR(\tauknu)}\xspace}					
\newcommand{\BFtautopinu}    {\ensuremath{\BR(\taupinu)}\xspace}
\newcommand{\BFtautohnunc}    {\ensuremath{\BR(\tauhnunc)}\xspace}
\newcommand{\BFtautoenu}    {\ensuremath{\BR(\tauenu)}\xspace}						
\newcommand{\BFtautomunu}    {\ensuremath{\BR(\taumunu)}\xspace}

\newcommand{\BRknu}   {\ensuremath{ 0.03882 \pm 0.00032 \pm 0.00057}\xspace}
\newcommand{\BRpinu}   {\ensuremath{0.5945 \pm 0.0014  \pm 0.0061}\xspace}

\newcommand{\BRmunu}   {\ensuremath{0.9796 \pm 0.0016  \pm 0.0036}\xspace}

\newcommand{\tauknulumi}             {\ensuremath{467\invfb}\xspace}

\newcommand{\gevccgevcc}{\ensuremath{{\mathrm{\,Ge\kern -0.1em V^2\!/}c^4}}\xspace}

\newcommand{\evcc}{\ensuremath{{\mathrm{\,e\kern -0.1em V\!/}c^2}}\xspace}

\makeatletter
\DeclareRobustCommand\bfseries{%
  \not@math@alphabet\bfseries\mathbf
  \fontseries\bfdefault\selectfont\boldmath}

\makeatother

\renewcommand{\babar}{\mbox{%
    \slshape B\kern-0.1em{\smaller A}\kern-0.1em
    B\kern-0.1em{\smaller A\kern-0.2em R}}\xspace}

\newcommand{\EE}[1]{\ensuremath{\cdot 10^{#1}}}

\newcommand{\CM}{CM\xspace}

\def\BR{{\ensuremath{\mathcal B}}\xspace}

\title{Measurements of |Vus| and Searches for
Violation of Lepton Universality and CPT in Tau Decays at \babar}

\ShortTitle{Measurements of |Vus| and Searches
for Violation of Lepton Universality and \CPT in Tau Decays at \babar}

\author{\speaker{Alberto Lusiani}%
        \thanks{Representing the \babar collaboration}\\
       Scuola Normale Superiore and INFN -- Pisa\\
       E-mail: \email{alberto.lusiani@pi.infn.it}}

\abstract{%
Using data collected with the \babar detector at PEP-II at SLAC, we
report on several tau lepton measurements that are used to determine
the modulus of the Cabibbo-Kobayashi-Maskawa matrix element \Vus, and
to check the Standard Model predictions of lepton universality and \CPT
conservation.
}

\FullConference{35th International Conference of High Energy Physics\\
                July 22-28, 2010\\
                Paris, France}

\begin{document}

\section{One-prong tau decay branching fractions}
\label{sec:oneprong}

A recent publication by the \babar collaboration reports on
precise measurements of the 1-prong tau decay branching
fractions~\cite{Aubert:2009qj}, using
\tauknulumi of \epem annihilation data at the \FourS peak,
produced at PEP-II (SLAC).

At the B-factories, precision tau branching fractions measurements are
limited by systematic uncertainties that are at present significantly
higher than the ones that were obtained at the LEP\emph{}
experiments\cite{Lusiani:2007cb}. Some of the largest systematic
uncertainties, such as the luminosity related ones, are common to
different channels, and cancel out when measuring ratios of branching
fractions. To best exploit systematics cancellations, the \babar
collaboration has measured in a single analysis four 1-prong tau
branching fractions: \BFtautoenu, \BFtautomunu, \BFtautopinu and
\BFtautoknu.

At the \FourS center-of-mass (\CM) energy, when
reconstructing tracks in the \CM frame, both tau leptons are boosted and
their decay products are well separated in two opposite hemispheres, which
can be conveniently defined using the reconstructed event thrust axis.
For all channels, signal candidates are selected by requiring an event
topology consisting in one prong in the signal side against three prongs
in the tag side. Tight particle identification is used in order to suppress
cross-feed background between the measured modes, achieving clean
selections.  Backgrounds, cross-feeds,
and the performance of particle identifications requirements are
carefully estimated using a variety of Monte Carlo and data control
samples.

The resulting measurements determine ratios of branching fractions where
the limiting systematics are determined by uncertainties on the particle
identification signal efficiencies and background suppressions, and on
uncertainties on background contaminations from channels other than the
ones that were simultaneously analysed. Systematics from luminosity,
from the $e^+e^-\to \tau^+\tau^-$ cross-section and from the branching
fraction and efficiency of modes on the remaining tau lepton into a
3-prong final state all cancel at first order.
Furthermore, using the electron branching fraction world
average~\cite{Amsler:2008zzb} as normalization, the \babar analysis
determines precise muon, pion and kaon branching fractions, 
which are comparable to the 2008 world averages~\cite{Amsler:2008zzb} 
for \BFtautomunu and \BFtautopinu and about four times more precise
for \BFtautoknu.
The measured ratios of branching fractions are
$\BFtautomunu / \BFtautoenu = \BRmunu$,
$\BFtautopinu / \BFtautoenu = \BRpinu$,
$\BFtautoknu$ $/$ $\BFtautoenu = \BRknu$.
The paper also quotes the full covariance matrix.

\section{Improvements on Lepton Universality Tests}

The above mentioned \babar publication~\cite{Aubert:2009qj} uses the
precise ratios of branching fractions to test lepton universality
for the weak charged current couplings.
A test of $\mu-e$ universality is obtained from the ratio
$\BFtautomunu/\BFtautoenu$, $(g_\mu/g_e)^2 = \gmgevalue$, which 
in combination with the world average~\cite{Amsler:2008zzb} yields
$(g_\mu/g_e)^2_\tau = 1.0018\pm0.0014$, which is
consistent with the Standard Model (SM) prediction and with the same
universality test that is obtained from pion
decays,
$(g_\mu/g_e)^2_\pi = 1.0021\pm0.0015$~\cite{Amsler:2008zzb,Cirigliano:2007ga}.
Using the precise ratios $\BFtautohnunc/\BFtautomunu$ with $h=\pi,K$ and
other well known quantities such as the tau, pion and kaon masses
and lifetimes~\cite{Amsler:2008zzb},
muon-tau universality tests are determined using respectively the pion
channel, the kaon channel, and both of them in combination:
$(g_\tau/g_\mu)^2_\pi = \gtgmpvalue$, 
$(g_\tau/g_\mu)^2_K = \gtgmkvalue$,
$(g_\tau/g_\mu)^2_h = \gtgmhvalue$.  The last value is 2.8$\sigma$ below
the SM expectation and within 2$\sigma$ of the world
average~\cite{Amsler:2008zzb}.

\section{Measurement of $\BR(\tau^-\to\KS\pim\piz\nut)$}

The \babar collaboration has produced a preliminary measurement of
$\BR(\tau^-\to\KS\pim\piz\nut)$~\cite{Paramesvaran:2009ec}.
Candidate events are required to have, in the \CM system, one hemisphere
containing one single prong identified as an electron or muon, and the
opposite hemisphere containing a \pim, a \pip\pim pair compatible with a
\KS decay, and a reconstructed \piz from two detected photons.
This channel is affected by large backgrounds from the processes where
the signal tau decays into $\KS\pim\nut$ and $\KS\pim\KL\nut$, however it is
possible to select candidates with relative high purity by requiring
that the \piz energy in the \CM system is larger than $1.2\gev$.
In so doing, one selects events in a restricted region of the phase
space, therefore it is crucial to have an accurate simulation of the
data in order to precisely estimate the efficiency of the selection.
Therefore \babar has studied the invariant mass distributions of all
couples of the final state particles (\KS, \pim, \piz) in the pure
selected sample of candidates and has tuned the Monte
Carlo generator to the data.
\iffalse
The measured branching fraction is
\begin{align*}
  \mathcal{B}(\tau^{-}\rightarrow \bar{K^{0}}\pi^{-}\pi^{0}\nu_{\tau}) = 
  \frac{1}{2N_{\tau\tau}} \frac{N_{\text{data}} -
    N_{\text{bkg}}}{\epsilon} =  0.342 \pm 0.006 \, \stat \pm 0.015 \, \syst,
  \label{eq:BR}
\end{align*}
where $N_{\text{data}}$ is the number of the selected candidates,
$N_{\text{bkg}}$ is the amount of background estimated with simulated events, 
$N_{\tau\tau}$ is the number of tau pairs events estimated using the
integrated luminosity and the $e^+e^-\to \tau^+\tau^-$ cross-section,
$\epsilon$ is the selection efficiency estimated with simulated events.
The largest systematic contribution comes from the uncertainty on the
\piz reconstruction efficiency, conservatively estimated at 3\%.
\else
The measured branching fraction is
$\mathcal{B}(\tau^{-}\rightarrow \bar{K^{0}}\pi^{-}\pi^{0}\nu_{\tau})
= 0.342 \pm 0.006 \, \stat \pm 0.015 \, \syst.$
\fi

\section{Determination of \Vus}
\newcommand{\Vudraw}{\ensuremath{V_{ud}}\xspace}
\newcommand{\Vusraw}{\ensuremath{V_{us}}\xspace}
\newcommand{\deltaRtauth}{\ensuremath{\delta R_{\tau,\text{SU3 breaking}}}\xspace} 

Several experimental measurements can be combined to determine the
modulus of the Cabibbo-Kobayashi-Maskawa matrix element \Vus, as
summarized by Table~\ref{tab:vusmethods}.  The most precise
determinations come from precise kaon measurements and are presently
limited by QCD theory systematic uncertainties.

\begin{table}[tb]\smaller
  \begin{center}
  \begin{tabular}{ll@{$\null=\null$}lcc}\toprule
    
    method & \multicolumn{2}{l}{illustrative formula} &
    $\Delta_{\text{th}}$ & Ref. \\
    \noalign{\smallskip}\midrule\noalign{\smallskip}
    
    $K_{\ell 3}$ & $\Gamma(K_{\ell 3})$ &
    $\left[\Vus {f_+(0)}\right]^2
    \times \left<\text{well known constants}\right>$ &
    0.58\% & \cite{Boyle:2010bh} \\
    \noalign{\smallskip}\midrule\noalign{\smallskip}
    
    $K_{\ell 2}$\enskip ($K/\pi$) &
    $\displaystyle\frac{\Gamma(K_{\ell 2})}{\Gamma(\pi_{\ell 2})}$ &
    $\displaystyle\left|\frac{\Vusraw}{\Vudraw}\right|^2
    {\frac{f_K^2}{f_\pi^2}}
    \times \left<\text{well known constants}\right>$ &
    0.50\% & \cite{Antonelli:2010yf} \\
    \noalign{\smallskip}\midrule\noalign{\smallskip}
    
    tau $K/\pi$ &
    $\displaystyle\frac{\Gamma(\tau\to K\nu)}{\Gamma(\tau\to\pi\nu)}$ &
    $\displaystyle\left|\frac{\Vusraw}{\Vudraw}\right|^2
    {\frac{f_K^2}{f_\pi^2}}
    \times \left<\text{well known constants}\right>$ &
    0.50\% & \cite{Antonelli:2010yf} \\
    \noalign{\smallskip}\midrule\noalign{\smallskip}
    
    $\tau\to K\nu$ exclusive &
    $\displaystyle\BR(\tauknu)$ &
    $\displaystyle\frac{G^2_F {f^2_K} \Vus^2 m^3_{\tau} \tau_{\tau}}{16\pi\hbar}
    \left (1 - \frac{m_K^2}{m_\tau^2} \right )^2 S_{EW}$ &
    1.27\%  & \cite{Follana:2007uv} \\
    \noalign{\smallskip}\midrule\noalign{\smallskip}
    
    $\tau\to X_s\nu$ inclusive &
    $\displaystyle\frac{R_{\tau,s}}{\Vus^2}$ &
    $\displaystyle\frac{R_{\tau,V+A}}{\Vud^2} - {\deltaRtauth}$ &
    0.23\%-0.47\% & \cite{Gamiz:2006xx,Gamiz:2007qs} \\
    \noalign{\smallskip}\midrule\noalign{\smallskip}
    
    unitarity &
    $\Vus^2$ & $1 - \Vud^2 - \Vub^2$ \\ \bottomrule
  \end{tabular}
  \caption{%
    Combinations of experimental measurements to compute \Vus:
    $\Delta_{\text{th}}$ identifies the theoretical uncertainties due to
    QCD as determined by the reference(s) on the right.
    $R_{\tau,s} = \Gamma(\tau\to\text{strange hadronic final states})/\Gamma(\tau\to e\nu\nub)$,
    $R_{\tau,V+A} = \Gamma(\tau\to\text{non-strange hadronic final states})/\Gamma(\tau\to e\nu\nub)$.
  }
  \label{tab:vusmethods}
  \end{center}
\end{table}

The above mentioned \babar publication~\cite{Aubert:2009qj}  reports two
\Vus determinations directly obtained from the measured branching
fractions, one using the ratio of kaon to pion branching fractions,
i.e.\ the ``tau $K/\pi$'' method in Table~\ref{tab:vusmethods} ($\Vus =
\Vusvalue$, using the \Vud measurement in Ref.~\cite{Hardy:2008gy}),
and a second one using the  ``$\tau\to K\nu$ exclusive'' method,
$\Vus= 0.2193 \pm 0.0032$.

The ratio $\Vus/\Vud$ can be determined from the ratio of the decay
widths of the tau to strange and non-strange hadronic final states, up
to a $SU(3)_{\text{flavor}}$ breaking correction determined by the $s$ quark mass
and QCD~\cite{Gamiz:2006xx}. Using the ratios
$R_{\tau,s} = \Gamma(\tau\to\text{strange hadronic final
  states}/\Gamma(\tau\to e\nu\nub)$ and
$R_{\tau,V+A} = \Gamma(\tau\to\text{non-strange hadronic final
  states}/\Gamma(\tau\to e\nu\nub)$,
the following relation holds, according to the ``$\tau\to X_s\nu$
inclusive'' method in Table~\ref{tab:vusmethods}:
\begin{align*}
  \Vus^2 = R_{\tau,s} / \left(\frac{R_{\tau,V+A}}{\Vud^2} - {\deltaRtauth}\right)
\end{align*}

To determine $R_{\tau,s}$ and $R_{\tau,V+A}$, we use all
tau branching fraction measurements reported in
Ref.~\cite{Amsler:2008zzb}, including
the 2009 partial update for the 2010 edition (PDG2009) and all the
available more recent measurements from the B-factories Belle and
\babar, including preliminary ones, as detailed in the HFAG 2010
update~\cite{TheHeavyFlavorAveragingGroup:2010qj}. We average all
measurements, taking into account correlations for the B-factories
measurements, in a single global unconstrained fit, with a
minimum \chisq fit. Since there is a large discrepancy between the
measurements of $\tau\to K^- K^+K^-\nu$ by \babar~\cite{Aubert:2007mh} and
Belle~\cite{Lee:2010tc}, a PDG-style error scale factor (5.44) has
been used.

We define the branching fraction of
the tau to a strange hadronic final state as described
in Table~\ref{tab:gamma110def} (which is slightly different from the
not-yet-finalized HFAG-tau
definition~\cite{TheHeavyFlavorAveragingGroup:2010qj}).

\begin{table}[tb]\smaller
  \begin{center}
    \begin{tabular}{lll}\toprule
      hadronic system in $\tau\to X_s \nu$ & fitted branching fraction \\
      \midrule
      $K^-$                             & $0.696\pm 0.010$ \\
      $K^-\pi^0$                       	& $0.431\pm 0.015$ \\
      $\bar{K}^0\pi^-$                 	& $0.827\pm 0.018$ \\
      $K^-\pi^0\pi^0$                  	& $0.060\pm 0.022$ \\
      $\bar{K}^0\pi^-\pi^0$            	& $0.348\pm 0.015$ \\
      $\bar{K}^0\pi^-\pi^0\pi^0$       	& $0.023\pm 0.023$ \\
      $K^-\pi^-\pi^+$                  	& $0.293\pm 0.007$ \\
      $K^-\pi^-\pi^+\pi^0$             	& $0.077\pm 0.012$ \\ 
      $K^-\pi^0\pi^0\pi^0$		& $0.040\pm 0.022$ \\ 
      $\bar{K}^0 h^-h^+h^-$	        & $0.023\pm 0.020$ \\ 
      $K^-\eta$                         & $0.016\pm 0.001$ \\ 
      $K^-\pi^0\eta$                    & $0.0048\pm 0.0012$ \\ 
      $\bar{K}^0\pi^-\eta$              & $0.0094\pm 0.0015$ \\ 
      $K^-K^+K^-$                       & $0.0023\pm 0.0008$ ($S=5.44$)      \\
      $K^-K^0\bar{K}^0$\quad from $K^-K^+K^-\cdot\frac{\phi\to K^0\bar{K}^0}{\phi\to K^+K^-}$
      & $0.0016\pm 0.0006$ ($S=5.44$)      \\ 
      \midrule
      $\tau\to X_s \nu$               & $2.854 \pm 0.052$ \\ 
      \bottomrule
    \end{tabular}
    \caption{%
      List of modes used to compute $\BR(\tau\to X_s \nu)$. Since the
      $\tau^-\to K^-K^+K^-\nut$ mode appears to be saturated by the
      $\tau^-\to K^-\phi\nut$ mode, we use the $\tau^-\to K^- K^+K^-\nu$
      measurement to estimate the $\tau^-\to K^- K^0K^0\nu$ mode using the
      $\phi$ branching fractions to $K^+K^-$ and $K^0K^0$. An error
      scale factor $S=5.44$ has been used to inflate the errors of the
      inconsistent \babar and Belle measurements for $\tau\to K^- K^+K^-\nu$.}
    \label{tab:gamma110def}
  \end{center}
\end{table}

\begin{figure}[tb]
  \begin{center}
    \begin{tabular}{@{\hspace*{0ex}}c@{}}
      {\fboxsep=2pt\fbox{%
          \begin{overpic}[trim=0 0 0 0,width=0.45\linewidth,clip]{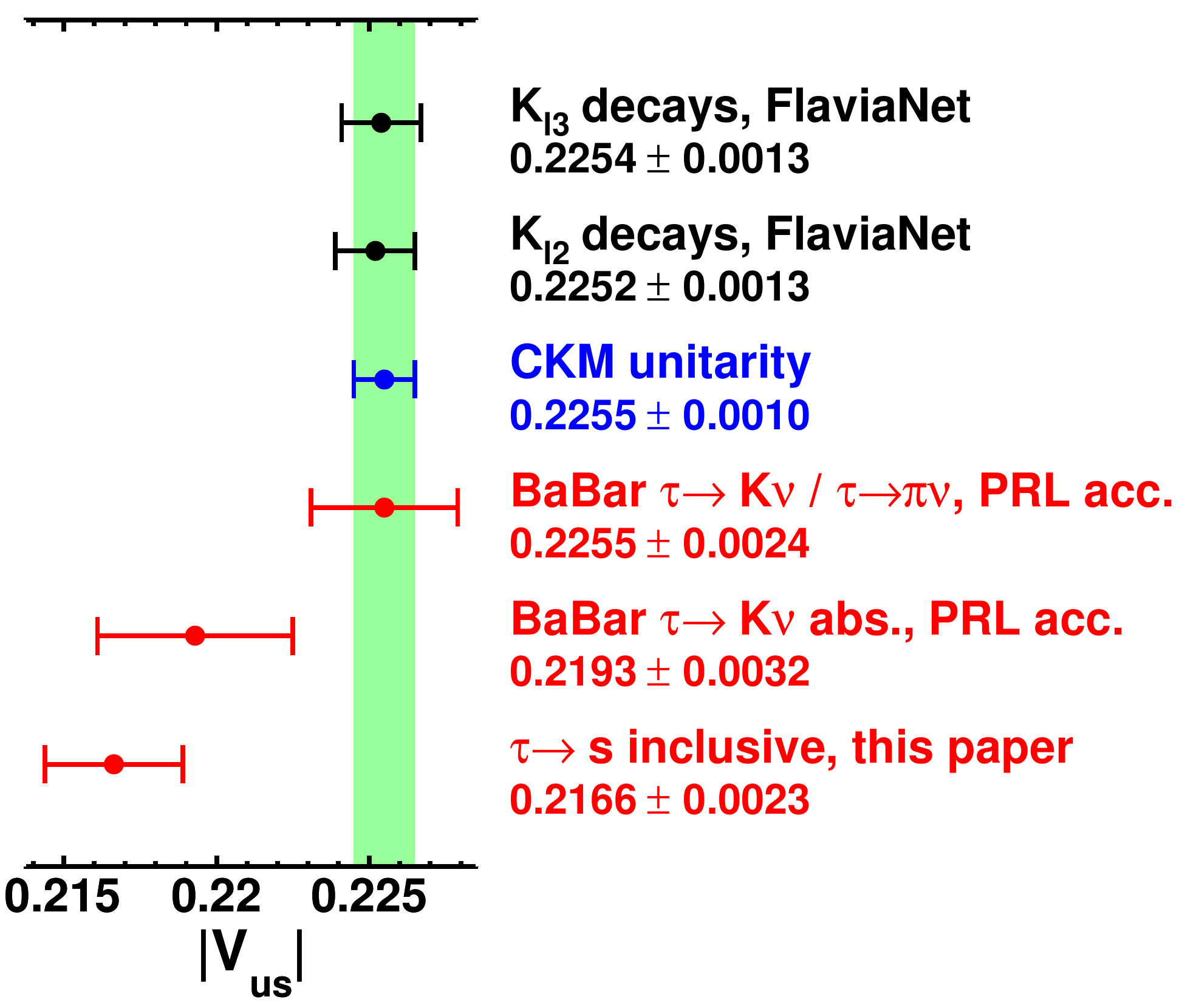}
          \end{overpic}}}
    \end{tabular}
    \caption{%
      Determinations of \Vus. The FlaviaNet numbers come from
      Ref.~\cite{Antonelli:2010yf}, the sources of the other data are
      mentioned in the text.}
    \label{fig:Vus}
  \end{center}
\end{figure}

In the literature, e.g.\ Refs.~\cite{Gamiz:2006xx,Gamiz:2007qs}, 
$R_{\tau,s}$ is obtained by summing all strange decay modes, while 
$R_{\tau,V+A}$ is obtained by difference using unitarity for sum
of the tau branching fractions, i.e.\ $R_{\tau,V+A}
= R_{\tau,h} - R_{\tau,s}$ with $R_{\tau,h} = (1-B_e -B_\mu)/B_e$,
where $B_e, B_\mu$ are the leptonic tau branching fractions computed
assuming SM lepton universality and using also the tau lifetime world
average~\cite{Davier:2005xq}.
This method implicitly assigns all possibly not-yet-measured or
neglected hadronic tau branching fractions to $R_{\tau,V+A}$.
To prevent that, we rather determine $R_{\tau,V+A}$ by summing up all
non-strange modes. We obtain $\Vus = 0.2166 \pm 0.0023$, which is
$3.57\sigma$ away from $\Vus = 0.2255 \pm 0.0010$ obtained from CKM
unitarity using Ref.~\cite{Hardy:2008gy}. We still use the improved
$B_e$ assuming SM lepton universality to compute $R_{\tau,V+A}$ and
$R_{\tau,s}$. Figure~\ref{fig:Vus} reports comparison with other \Vus
determinations.
When using the literature definition for $R_{\tau,V+A}$,
there are no significant changes: $\Vus = 0.2163 \pm 0.0023$,
$3.66\sigma$ away from unitarity.  By inflating measurement
errors in the global fit with a procedure similar to the one used by
PDG for error scale factors, we obtain $\Vus = 0.2166 \pm 0.0025$,
corresponding to a slightly reduced discrepancy of $3.29\sigma$.

\section{Test of CPT conservation from tau mass measurements}

The \babar collaboration has recently published a tau mass
measurement~\cite{Aubert:2009ra} obtained by fitting the end-point of
the reconstructed minimum tau mass for $\tau\to 3\pi\nu$ candidates.
The measurement is robust against background contamination, is
statistically precise but is ultimately systematically limited by the
calibration of the reconstructed momentum scale, remaining less
competitive with respect to measurements at the $\tau^+\tau^-$
production threshold that use resonant beam depolarization to
measure the beam energy.  However, unlike threshold experiments,
\babar can measure the mass of positive and negative tau leptons
separately, providing a competitive test of \CPT conservation.
The tau mass is measured to be $m_\tau = 1776.68 \pm 0.12\,\stat \pm
0.41\,\syst\mev$ and the mass difference is
$\left(m_{\tau+} - m_{\tau-}\right)/m_{\text{average}} =
-3.4 \pm 1.3\,(\text{stat.}) \pm 0.3\,(\text{syst.})\EE{-4}$.
There is no evidence of \CPT violation as in the similar published
measurement by Belle~\cite{Abe:2006vf}.

\bibliographystyle{JHEP}
\bibliography{%
  ichep10-bib,
  pub-a-pre90,
  pub-b-90-94,
  pub-c-95-99,
  pub-d-00-04,
  pub-e-05-09,
  pub-extra,
  tau-lepton
}

\end{document}